# Giant stellar arcs in the Large Magellanic Cloud: a possible link with past activity of the Milky Way nucleus

Yuri N. Efremov*

*Sternberg Astronomical Institute of the Lomonosov Moscow State University, Moscow 119992, Russia*

**ABSTRACT**

The origin of the giant stellar arcs in the LMC remains a controversial issue, discussed since 1966. No other star/cluster arc is so perfect a segment of a circle; moreover, there is another similar arc near-by. Many hypotheses were advanced to explain these arcs, and all but one of these were disproved. It was proposed in 2004 that origin of these arcs was due to the bow shock from the jet, which is intermittently fired by the Milky Way nucleus – and during the last episode of its activity the jet was pointed to the LMC. Quite recently evidence for such a jet has really appeared. We suppose it was once energetic enough to trigger star formation in the LMC, and if the jet opening angle was about 2°, it could push out HI gas from the region of about 2 kpc in size, forming a cavity LMC4, - but also squeezed two dense clouds, which occurred in the same area, causing the formation of stars along their surfaces facing the core of the MW. In result, spherical segments of the stellar shells might arise, visible now as the arcs of Quadrant and Sextant, the apices of which point to the center of the MW. This orientation of both arcs can be the key to unlocking their origin. Here we give data which confirm the above hypothesis, amongst which are radial velocities of stars inside and outside the larger one of the LMC arcs. The probability is low that a jet from an AGN points to a nearby galaxy and triggers star formation there, but a few other examples are now known or suspected.

## 1 INTRODUCTION

The origin of two large arcs of young stars and clusters, which are near each other in the north-east outskirts of the Large Magellanic Cloud, is an important unsolved problem. These arcs are parts of perfect circles, the stars within each arc being practically coeval. The smaller arc is essentially younger than the larger one and surrounded by HII regions. The large arc was called "Quadrant" by Efremov & Elmegreen (1998), and the smaller one - "Sextant".. These two arcs are unique objects. They are segments of the regular circles with radii about 300 and 200 pc, ages of stars and clusters there being within 12 - 20 and 4 -7 Myr. Apices of both arcs pointed to the Milky Way center. Two or even three more arcs might be suspected in the LMC4 region (Hodge 1967), but they are far from the regularity and richness of Quadrant and Sextant arcs and are likely by chance configurations. The Quadrant arc was first noted by Westerlund and Mathewson (1966), who wrongly identified it as Shapley's "Constellation III". In fact, McKibben Nail & Shapley (1953), introducing the term "constellation", noted the NGC 1974 cluster as the identifier of Constellation III, - and this cluster in fact is located within the Sextant arc (Fig. 1).

----------------------------------

*E-mail: efremovn@yandex.ru

Bok et al. (1962) were the first to study it, noting by the way the arc-like appearance of this association, which includes, apart from NGC 1974, the clusters NGC 1955 and NGC 1968. The Quadrant arc (known also as OB-association LH77) is inside the HI hole LMC4; this was a reason for Westerlund & Mathewson's (1966) suggestion that both these structures were formed in result of a Super-Supernova outburst. They have not noted the smaller Sextant arc which is outside the LMC4 "superbubble", but soon Hodge (1967) briefly described both these arcs, as well as a somewhat similar formation in the spiral galaxy NGC 6946. Prof. Hodge informed us that this was the only such structure he found in systematic searches for features resembling the stellar arcs in the LMC.

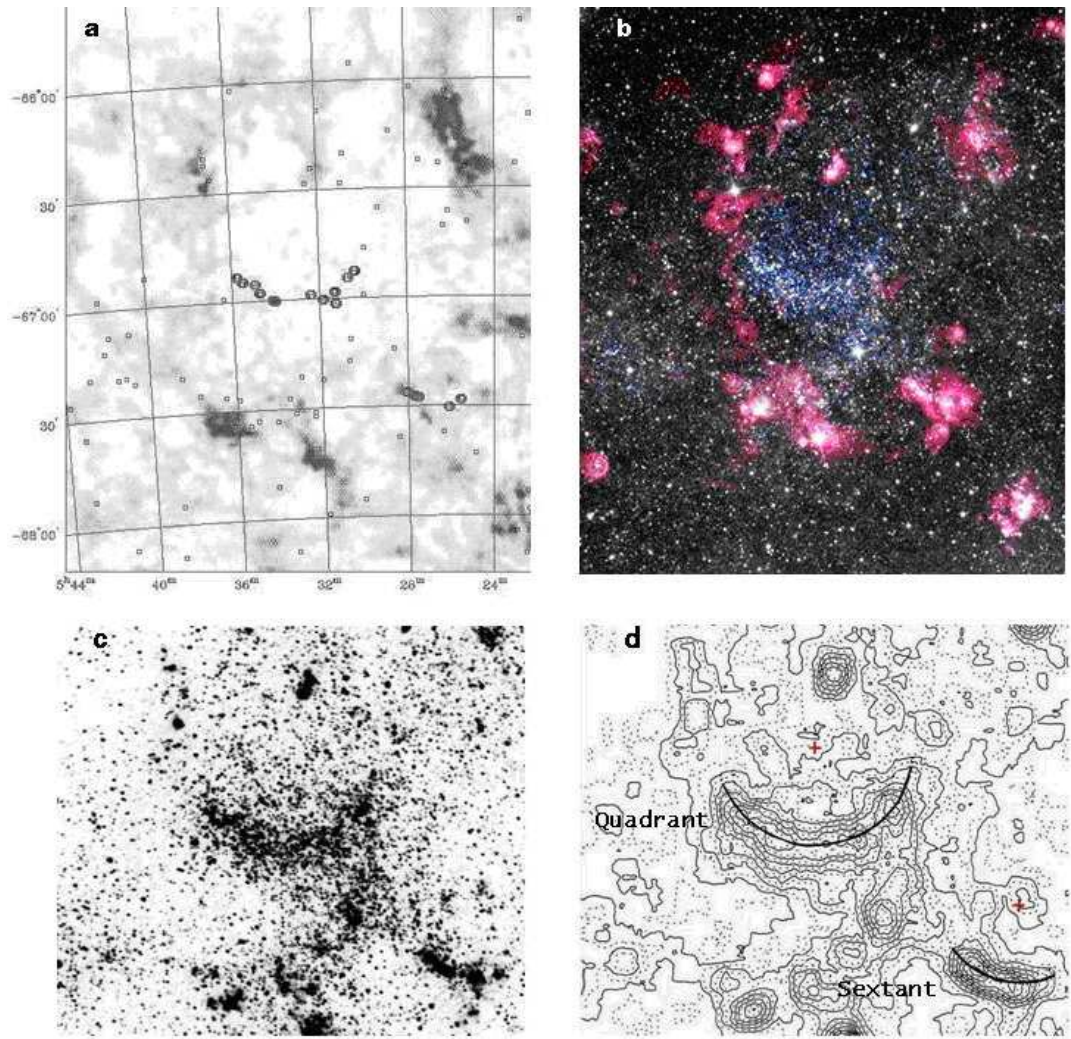

**Figure 1**. The LMC4 region in NE outskirts of the Large Magellanic Cloud.
(a) HI and star clusters in the LMC 4 region (courtesy of Jan Palous). Clusters within Qudrant and Sextant arcs are shown as circles. 1° = 900 pc.
(b) The same area with HII regions shown (http://apod.nasa.gov/apod/ap110426.html).
(c) The LMC4 region in image, digitized from the old photographic plate obtained at Boyden Observatory (courtesy of Paul Hodge).
(d) Isodensities from the preceding image, with segments of circles through Quadrant and Sextant added.

The regular circular shape of each arc is really striking, especially in the U band. Efremov & Elmegreen (1998) were the first to point out the problem of the origin of the whole system of the LMC arcs. The circular shapes of these two arcs were explained as a consequence of their formation through gas being swept by the pressure of O stars and supernovae possibly residing once at the arc centers.

However, later Efremov & Elmegreen (1999) pointed out the logical flaws of this hypothesis. It is unclear why such unusual features as these regular arcs should have been formed around quite inconspicuous clusters, while being absent around all other known much richer clusters. It seems impossible to understand, why both giant stellar arcs known in the LMC are located near each other.

As a source of energy, the superbursts of GRBs were suggested (Efremov, Elmegreen & Hodge 1998), and the side by side location of the LMC arcs was later explained by *ad hoc* hypothesis on their GRB progenitors escaping from their birthplace in the dense old cluster NGC 1978, which is inside the LMC4 superbubble (Efremov 2001). However there are a lot of dense massive clusters elsewhere, as well as GRBs, and yet no stellar arcs are known as regular as those two in the LMC.

The suggestion that these arcs were formed by some source of pressure located inside them seems to be untenable now. Here we consider the idea about their origin in two large and dense HI clouds, which was due to an outer source of pressure, located far from the LMC.

## 2 ORIGIN OF THE LMC STELLAR ARCS

We need to explain why the arcs are near each other and their orientations are the same, and to find how the prestellar gas structures were themselves formed. As the remarkable feature of both two arcs is their circularity (Fig. 1), it was natural to suggest they were produced from nearly uniform gas swept up by a central source of pressure (Efremov & Elmegreen 1998; Efremov, Ehlerova & Palous 1999). It is believed that such a pressure lead to expansion of HI shells/bubbles and gravitational instability along these shells lead to star formation around them.

This is indeed observed; probably the best case is the southern rim of the large (~2 kpc in size) the HI/dust cavity #107 in NGC 6946, along which we noted a number of HII regions/O-stars clumps, forming a giant, rather irregular arc (Efremov et al. 2011). However, such a mechanism cannot be applied to explain the LMC arcs. The Quadrant arc is near the center of LMC4 hole, while the Sextant is outside it; neither of these arcs is located along the LMC 4 edge. The borderlines of this superhole are quite irregular and there is no a visible shell around it (unlike to supershell #107 in NGC 6946); it may be even a superposition of a few smaller holes in the HI disk (Fig. 1a).

The strictly circular shapes of both arcs recalls the bow shock appearance of the leading edges of some galaxies, which is due to ram pressure, arising from a galaxy motion through the intergalactic medium within a cluster of galaxies. Evidence for ram pressure is well known, especially in a number of galaxies – members of the Virgo cluster, but also in some others. For example, segments of the perfect circles outline the HI edges of the Ho II irregular galaxy (Bureau & Carignan 2002) and of the spiral galaxy NGC 7421, the latter demonstrating also enhanced star formation along the leading edge (Ryder et al. 1997).

Physically the same mechanism may trigger star formation process owing to interaction of an energetic jet with a sufficiently dense gas medium. The jet pressure may lead to formation of a partial stellar shell, whose convex side turned to the jet source, if the target gas cloud is smaller than the jet cross-section at the distance of the former. The hypothesis based on these considerations was advanced - the LMC arcs might be formed by the bow shock pressure from a jet originating at the Milky Way nucleus during periods of its activity (Efremov 2004). This proposal was prompted by the observation that apices of both arcs point to the Milky Way center. (Direction to the 30 Dor region in the LMC is about the same, but no source of a jet energetic enough is expected there.)

At that time there was no data on such a jet, but it has been suspected already that the MW is a micro-AGN galaxy, a kind of faint Seyfert galaxy (Mezger et al. 1996), and that the Galactic center is driving bipolar, large scale winds into the halo every 10-15 Myr (Bland-Hawthorn & Cohen 2003). There exist now many new data on the intermittent (approximately each 10 Myr) activity of the Milky Way nucleus (e.g. Guo & Mathews 2012, Silk et al. 2012), and on the transient existence of jets from the Galactic nucleus. The final proof was given recently by Su & Finkbeiner (2012), which found Gamma-ray jets of some 10 kpc long each. The present day Southern jet does not point to the LMC, but it is known that the jets from active galactic nuclei are precessing and variable in intensity, at various time scales, up to many millions years. It might have once pointed to the LMC. At any rate, the recurrent activity of the Milky Way nucleus is now established for sure.

The best studied case of star formation triggered by a jet from another galaxy is the Minkowski object, the peculiar irregular galaxy, where a burst of star formation was triggered by the impact of the radio jet from the nucleus of NGC 541 galaxy, which is about 15 kpc distant First evidence for such triggering were given about simultaneously by Brodie, Bowyer & McCarthy (1985) and by van Breugel et al. (1985). This case now is well studied and the conclusion is the same (Croft et al. 2006). There are many data for much more distant objects and more power jets, suggested they are able to trigger star formation at distances of a hundred kpc from the parent nucleus. Also, it was found that strong ram pressure from the radio jet originating at the center of the galaxy NGC 3079 have blown HI out of the nearby galaxy NGC 3073 (Irwin et al. 1987).

It can be assumed now that the jet from the core of the Milky Way was once directed to the outskirts of the LMC and it was energetic enough to trigger star formation there. If the jet-cross section at the LMC distance was about 2 kpc (close to the size of LMC4), this means that the opening angle of the jet at the Galactic nucleus was about 1 - 2°, as observed in Cyg A. Further on, we have to assume that there two dense gas clouds occured, both of which had smaller diameter than the cross-section of the jet at a distance of the LMC. The jet could push out HI gas from the region of about 2 kpc in size, forming a cavity LMC4, but also squeezed two more dense clouds in the same area, causing the formation of stars along their surfaces facing the core of the MW, resulting in a perfectly spherical segments of the stellar shells, visible now as the arcs of Quadrant and Sextant, apices of which pointed to the center of the MW. This orientation of both arcs can be really the key to unlocking their origin.

The different ages of the LMC arcs conflict with the hypothesis of a unique outburst by the MW jet. One might however suggest that the jet was quite narrow and changed a bit its direction/activity during some 10 Myr or so. There is anyway more plausible explanation. Harris & Zaritsky (2008), using their vast photometric data, reconstructed the star-formation history of the "Constellation III" (in fact, within the 2.5 x 2.5 degree region, centered on the Quadrant arc) and found that star formation there was most active during two distinct epochs. It follows from their Fig. 4 that the highest star formation rate was 25 – 8 Myr ago within Quadrant arc and 13-8 Myr ago within the Sextant arc. The age difference between these two arcs is probably even larger than above estimation, as the Sextant arc hosts O-stars.

Harris & Zaritsky's (2008) conclusion was that "the prestellar gas is somehow pushed into these large-scale arc structures, without simultaneously triggering immediate and violent star formation throughout the structure. Rather, star formation proceeds in the arc according to the local physical conditions of the gas". The difference in these conditions in the parent clouds might well lead to the different average age of two star/cluster arcs, arisen along the cloud surfaces, turned to the MW core.

As concerns the LMC4 HI void, it might otherwise be a later result of pressure from the Quadrant stars, as Efremov & Elmegreen (1998) have suggested, - and was not formed by the jet pressure. There is no HI hole around the Sextant arc, probably because it lies outside the main HI layer of the LMC, like a situation which is observed in the 30 Dor region (van der Marel & Cioni 2001). The Quadrant and Sextant arcs are seen as segments of regular rings instead of ellipses appropriate to inclination of the LMC plane to the sky plane, which may be explained by the large thickness of the LMC gaseous disk and, respectively, of the stellar arcs. The arcs are partial thick shells, and not the segments of plane circles.

# 3 RADIAL VELOCITIES OF STARS IN THE LMC4 REGION

Ram pressure, leading to star formation in the gas cloud, also accelerates this cloud and endows newly formed stars with momentum (Gaibler et al. 2012). In our case, the radial velocities of the LMC stars should be larger (more positive) for stars formed in gas influenced by the bow shock from the MW jet. Courtesy of Mary Kontizas and Russell Cannon, who have supplied the respective data, permitted to compare the radial velocities for the high luminosity stars in the LMC4 region - inside and outside the Quadrant arc.

As Fig. 2 and 3 demonstrate, within the Quadrant arc (subregion *d*) the velocities are more positive, indeed; moreover, their dispersion is twice lower than inside other three subregions. Thus, the bulk motion of stars within the Quadrant confirms, or at least does not contradict the suggestion that the LMC stellar arcs originated in interaction of the gas clouds with an energetic jet from the MW core.

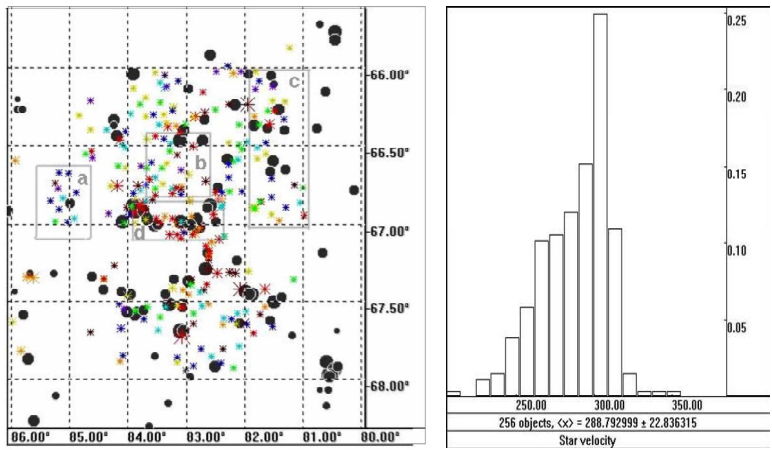

**Figure 2.** The radial velocities of the high luminosity stars in the LMC4 region. The map and Vr distribution are shown; the redder color in the map, the larger Vr. Black circles are star clusters.

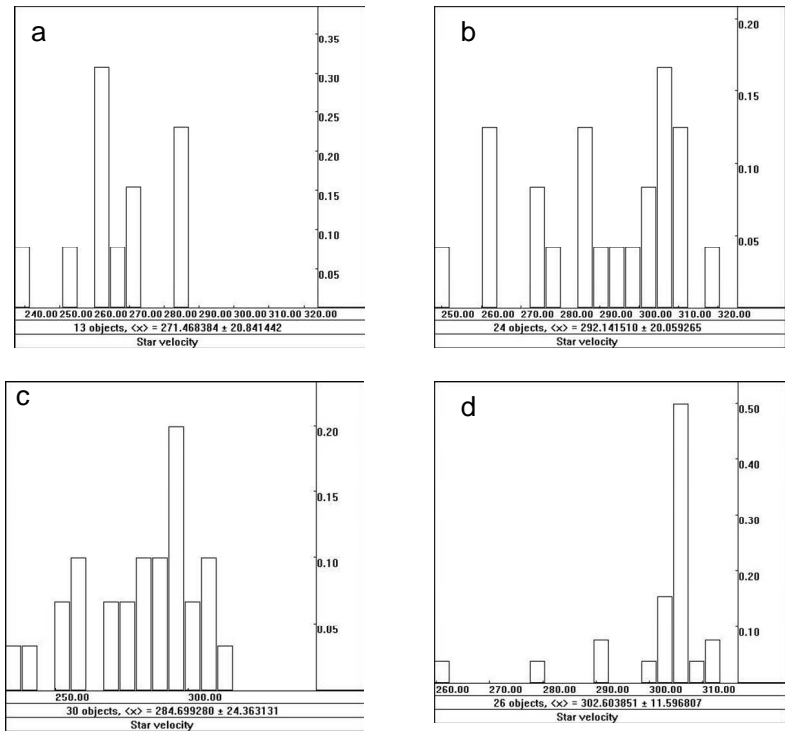

**Figure 3.** Distributions of Vr inside the subregions, designated in Fig. 2. The average Vr values and their dispersions (km/s) are shown.

## 4 CONCLUSIONS

The Large Magellanic Cloud stellar arcs are unique objects. Considering the orientation of the arcs, the evidence of the contemporary activity of the Milky Way nucleus (which is rather low now but surely was high in the past), and the circular shape of arcs, resembling the edges of HI haloes of galaxies, shaped by ram pressure, we suggest that the arcs were formed by bow shock from the jet fired by the Milky Way nucleus a dozen Myr ago. This suggestion is quite special, but is plausible nonetheless. Changes of orientation and power are quite common for the jets from AGNs, and there exist more examples of star formation triggered by such jets. Jet-induced star formation in gas-rich galaxies is becoming accepted as a rather common phenomenon (Gaibler et al. 2012). Star formation, triggered by a jet from the active nucleus of another galaxy is rare event, but it may occur. Thus, Pashchenko & Vitrishchak (2010) estimated the frequency of such events to be of the order of several percent of all close (with separations of the order of several tens of kpc) pairs containing a radio AGN. We are seemingly forced to accept such a special scenario for the origin of the LMC arcs because other explanations of their properties are even less probable.

## ACKNOWLEDGEMENTS

I am thankful to Paul Hodge, Bruce Elmegreen, and Jan Palous, with whom the previous attempts to disclose the enigma of the LMC stellar arcs were undertaken. Special thanks are due to Mary Kontizas and Russell Cannon who provided me with the radial velocities of the high luminosity stars in the LMC4 region. I am indebted to Eugene Efremov for help in construction of Figure 2.
Many thanks are due to the reviewer James Binney, whose corrections of the text and suggestions of more exact formulations helped to present more clearly the hypothesis on the LMC arcs origin.
This work was supported by the RFFI grant # 12-02-00827-a.